\documentclass{article}

\usepackage{arxiv}

\usepackage[utf8]{inputenc} % allow utf-8 input
\usepackage[T1]{fontenc}    % use 8-bit T1 fonts
\usepackage{hyperref}       % hyperlinks
\usepackage{url}            % simple URL typesetting
\usepackage{booktabs}       % professional-quality tables
\usepackage{amsfonts}       % blackboard math symbols
\usepackage{nicefrac}       % compact symbols for 1/2, etc.
\usepackage{microtype}      % microtypography
\usepackage{lipsum}
\usepackage{cleveref}
\usepackage{graphicx}
\usepackage{epsfig}
\usepackage{subfig}
\usepackage{natbib}
\title{Mid-IR non-volatile silicon photonic switches using nanoscale Ge\textsubscript{2}Sb\textsubscript{2}Te\textsubscript{5} embedded in SOI waveguide}

\author{
  Nadir Ali \\
  Department of Physics\\
  Indian Institute of Technology Roorkee\\
  Roorkee, India 247667 \\
  \texttt{nali@ph.iitr.ac.in} \\
   \And
 Rajesh Kumar \\
 Department of Physics\\
  Indian Institute of Technology Roorkee\\
  Roorkee, India 247667\\
  \texttt{rajeshfph@iitr.ac.in} \\
}

\begin{document}
\maketitle

\begin{abstract}
We propose and numerically analyze the hybrid Si-Ge\textsubscript{2}Sb\textsubscript{2}Te\textsubscript{5} strip waveguide switches for Mid-IR wavelength of 2.1 $\mu$m. The switches investigated are one input-one output (on-off) type and one input-two outputs (directional coupler) type. The reversible transition between the switch states is achieved by inducing the phase transition from crystalline to amorphous and vice-versa by application of voltage pulses. The approach of embedding the nanoscale active material Ge\textsubscript{2}Sb\textsubscript{2}Te\textsubscript{5} within the Si waveguide is taken to enhance the interaction of light with the active region of the switches. The dimensions of the active regions of the switches are optimized to achieve low insertion loss, low switching energy and high extinction ratio. In case of one input-one output switch, an extinction ratio of 33.79 dB along with an extremely low insertion loss of 0.52 dB is achieved using optimum Ge\textsubscript{2}Sb\textsubscript{2}Te\textsubscript{5} length of only 0.92 $\mu$m. For one input-two outputs switch, an extinction ratio of 10.33 dB and 5.23 dB is obtained in cross and bar state respectively using an active length of 52 $\mu$m. These values of extinction ratio, which are otherwise 18.59 dB and 8.33 dB respectively, are due to the necessity of doping the Si beneath the Ge\textsubscript{2}Sb\textsubscript{2}Te\textsubscript{5} to facilitate the electrical conduction needed for Joule heating. A suitable gap of 100 nm is kept between the active and passive arm of the directional coupler switch. The electro-thermal co-simulations confirm that phase change occurs in whole of the Ge\textsubscript{2}Sb\textsubscript{2}Te\textsubscript{5} region in both types of switches.
\end{abstract}
% keywords can be removed
\section{Introduction}
The internet data traffic over the fiber optic network has been increasing at an exponential rate and rapidly approaching the projected ultimate data transmission capacity limits, raising concerns over a future capacity crunch. Thus, a new non-conventional physical technology will soon be indispensable \cite{b1}.\\
\begin{figure}[t]
	\centering
		\includegraphics[scale=.5]{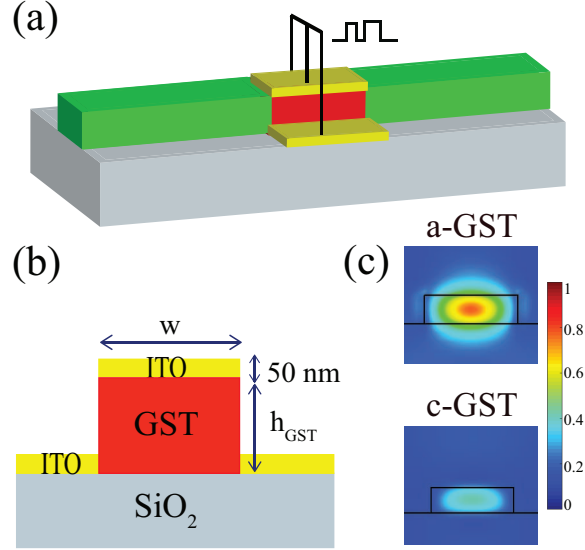}
	\caption{One input-one output switch design using fully etched Si waveguide. (a) Schematic of hybrid Si-GST 1 $\times$ 1 waveguide switch. (b) Cross-section of the active region of switch; phase change material GST substituted in fully etched silicon waveguide. (c) Mode profile in the ON (a-GST) and the OFF (c-GST) state of switch.}
	\label{fig1}
\end{figure}
A viable solution is to open a new optical communication wavelengths window around 2 $\mu$m region \cite{b2}. The main motivation behind this is to exploit the recently developed hollow core photonic band gap fiber (HC-PBGF) and Thulium doped fiber amplifier. The HC-PBGFs guides light in the air core, which minimize the Rayleigh scattering loss by having high wavelength of operation as Rayleigh scattering has $\lambda^{-4}$ dependence, and in fact for such fibers minimum loss window falls around 2 $\mu$m region. In addition, HC-PBGFs exhibit low nonlinearity and low latency as compared to the conventional fibers \cite{b3,b4}. Combining HC-PBGF with Thulium doped amplifier, which has exceptionally high band width in this region \cite{b5}, projects an attractive future for the optical communication around 2 $\mu$m wavelength. However, a fully functional next-generation photonic system will require a full range of suitable components with higher integration, miniaturization and complex functionalities. The advantages of silicon (Si) over other materials for conventional optical communication wavelengths have been studied and stated in the literature, and this can be further extended to the 2 $\mu$m region. Si is transparent for the wavelength range of 1-7 $\mu$m \cite{b6}, and Si photonic components can be fabricated using mature complementary metal oxide semiconductor (CMOS) technology employed in manufacturing of the electronic components \cite{b7}. However, Si active photonic devices such as switches and modulators are bulky and consumes large power because of the modest tunability of Si and requirement of a continuous power source to maintain the bias \cite{b8}. Therefore, non-volatile switches that consume lesser power and have ultra-compact footprint are required to unleash the full potential of optical transmission around 2 $\mu$m wavelength.\\
\begin{figure}[b]
	\centering
		\includegraphics[scale=.5]{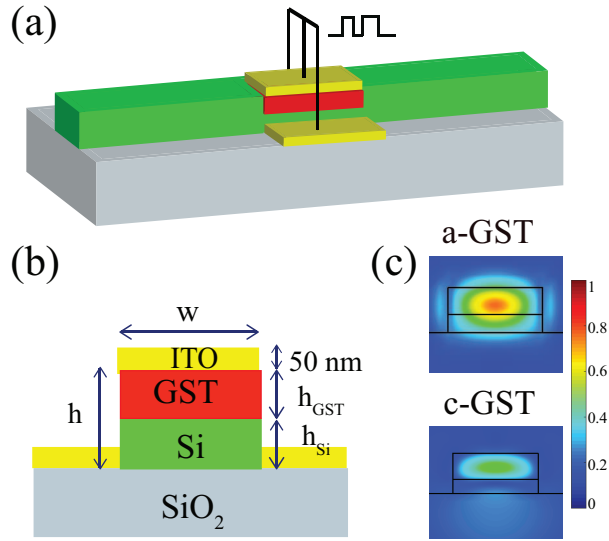}
	\caption{One input-one output switch design using partially etched Si waveguide. (a) Schematic of hybrid Si-GST 1 $\times$ 1 waveguide switch. (b) Cross-section of active region of switch; phase change material GST substituted in partially etched silicon waveguide. (c) Mode profile in the ON (a-GST) and the OFF (c-GST) state of switch.}
	\label{fig2}
\end{figure}
Recently, a well-known phase change material Ge\textsubscript{2}Sb\textsubscript{2}Te\textsubscript{5} (GST) has been utilized as an active material in non-volatile on-chip active devices on different photonic platforms \cite{b9}. The active operation in these devices is enabled because GST exhibits good phase change behavior among its amorphous (a-GST) and crystalline (c-GST) phases. The transition between two phases of GST is rapid, repeatable and results in a huge difference in optical and electronic properties. Because of high refractive index difference between the two phases of GST, active devices exhibit high on-off ratio and can have broad band operation with ultra-compact footprint \cite{b10}. Another interesting property of GST-based devices is the self-holding bistability i.e. no continuous power is required to maintain the device state \cite{b11}. The switching of GST from the a-GST to the c-GST and vice-versa is achieved by Joule heating GST material through the application of optical and electrical pulses \cite{b12,b13}. These transitions occurs with transition time of sub-nanoseconds \cite{b14}. Moreover, the fabrication of GST based devices is CMOS compatible.\\
\begin{figure}
	\centering
		\includegraphics[scale=.5]{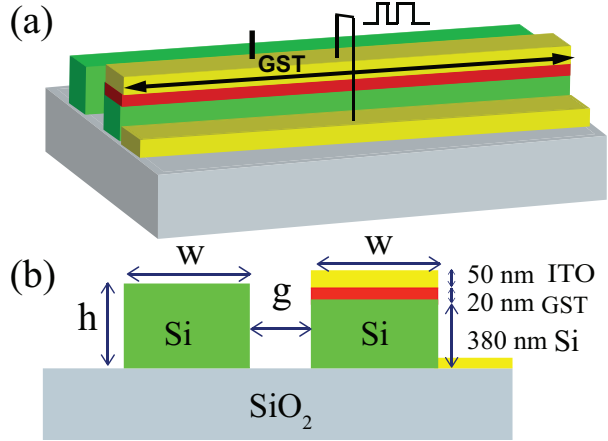}
	\caption{(a) Schematic of hybrid Si-GST 1 $\times$ 2 directional coupler switch: only the coupling waveguides of directional coulper are shown. (b) Cross-section of the coupling waveguides where a 20 nm thick GST substituted in the partially etched silicon waveguide.}
	\label{fig3}
\end{figure}
Previous studies reported chip-based GST experiments on switching and modulation for the optical communication 1.55 $\mu$m wavelength \cite{b15,b16,b17}. In most cases, the GST material was deposited on top surface of the waveguide and acts as a variable refractive index cladding. Here, we take another approach by embedding GST inside the etched silicon-on-insulator (SOI) waveguides. Such an approach improves the light-matter interaction and provide a high extinction ratio with small active volume \cite{b18}.\\
In this article, we report switches aimed for operation at 2.1 $\mu$m wavelength, using GST embedded within the SOI waveguide. The wavelength selection is inspired from the fact that the GST has lowest attenuation coefficient $k_{am}$ = 0.006 at 2.1 $\mu$m wavelength around 2 $\mu$m window \cite{b19}. The GST material is embedded within the etched Si waveguide to provide high overlap between the optical mode in the waveguide and the GST. Taking into consideration both insertion loss and extinction ratio, we find the optimized value of GST dimensions. For the phase transition of GST, we consider voltage triggered phase change for switching applications. Hence, electro-thermal co-simulations are performed for the optimized structure of switches.\\
The rest of the paper is organized as follows. In the second section, we presents the design of Si waveguide, geometries of switches, and the parameters used to analyze switches. In the third section, results obtained by performing electromagnetic and thermal simulations are described. Finally, the conclusions are presented in the fourth section.\\
%%%%%%%%%%%%%%%%%%%%%%%%%%%%%%%%%%%%%%%% Design And Simulations %%%%%%%%%%%%%%%%%%%%%%%%%%%%%%%%%%%%%%%%
%%%%%%%%%%%%%%%%%%%%%%%%%%%%%%%%%%%%%%%%%%%%%%%%%%%%%%%%%%%%%%%%%%%%%%%%%%%%%%%%%%%%%%%%%%%%%%%%%%%%%%%%
\section{Designs and simulations}
The switches designed here are hybrid Si-GST one input-one output (1 $\times$ 1) waveguide switch and one input-two outputs (1 $\times$ 2) directional coupler switch. The Si strip waveguide is designed in SOI platform with a 400 nm height Si layer and a 2 $\mu$m thick buried silicon dioxide substrate. For the single mode operation at 2.1 $\mu$m the cross section of waveguide is 800 nm $\times$ 400 nm (width $\times$ height). The approaches taken to design hybrid Si-GST switches are described below.
\subsection{1 $\times$ 1 waveguide switch}
In 1 $\times$ 1 waveguide switch, the output optical power through the Si waveguide is altered by changing the phase of GST (from a-GST to c-GST and vice-versa) embedded inside the Si waveguide. The optical mode is launched from the left side and output is measured from the right side of the waveguide shown in figure~\ref{fig1}. To achieve high overlap between the optical mode and the GST, first, we fully etched the Si waveguide in the middle and embedded GST in the trench as shown in  figure~\ref{fig1}(a). The indium-tin-oxide (ITO) electrodes are used to inject voltage pulses into GST to induce GST phase transition as shown in figure~\ref{fig1}(a). The ITO is chosen as electrode material as its deposition temperature is around 300 $^{\circ}$C. This temperature is CMOS compatible and ITO as electrode material can be introduced in CMOS process \cite{b20}. As figure~\ref{fig1}(b) shows, the height of embedded GST is denoted as $h_{GST}$ and the length of active region as $l_{GST}$ (not shown in figure~\ref{fig1}). The same figure also shows the electrode material ITO layers for top and bottom electrical contacts. Each of the ITO layer has height of 50 nm. To achieve high extinction ratio and low insertion loss, we varied GST dimensions $h_{GST}$ and $l_{GST}$. Because of different values of GST extinction coefficient, the optical mode suffers variable attenuation in two phases. In the amorphous phase, the mode experience smaller attenuation. While light is highly absorbed when GST is in the crystalline phase because it has higher value of extinction coefficient than amorphous phase. This can be seen from the mode profile inside the active region of hybrid Si-GST waveguide switch shown in figure~\ref{fig1}(c). Thus, the a-GST and the c-GST phases exhibits the ON and the OFF states of switch, respectively.\\
\begin{figure}
\centering
\subfloat{\includegraphics[scale=0.5]{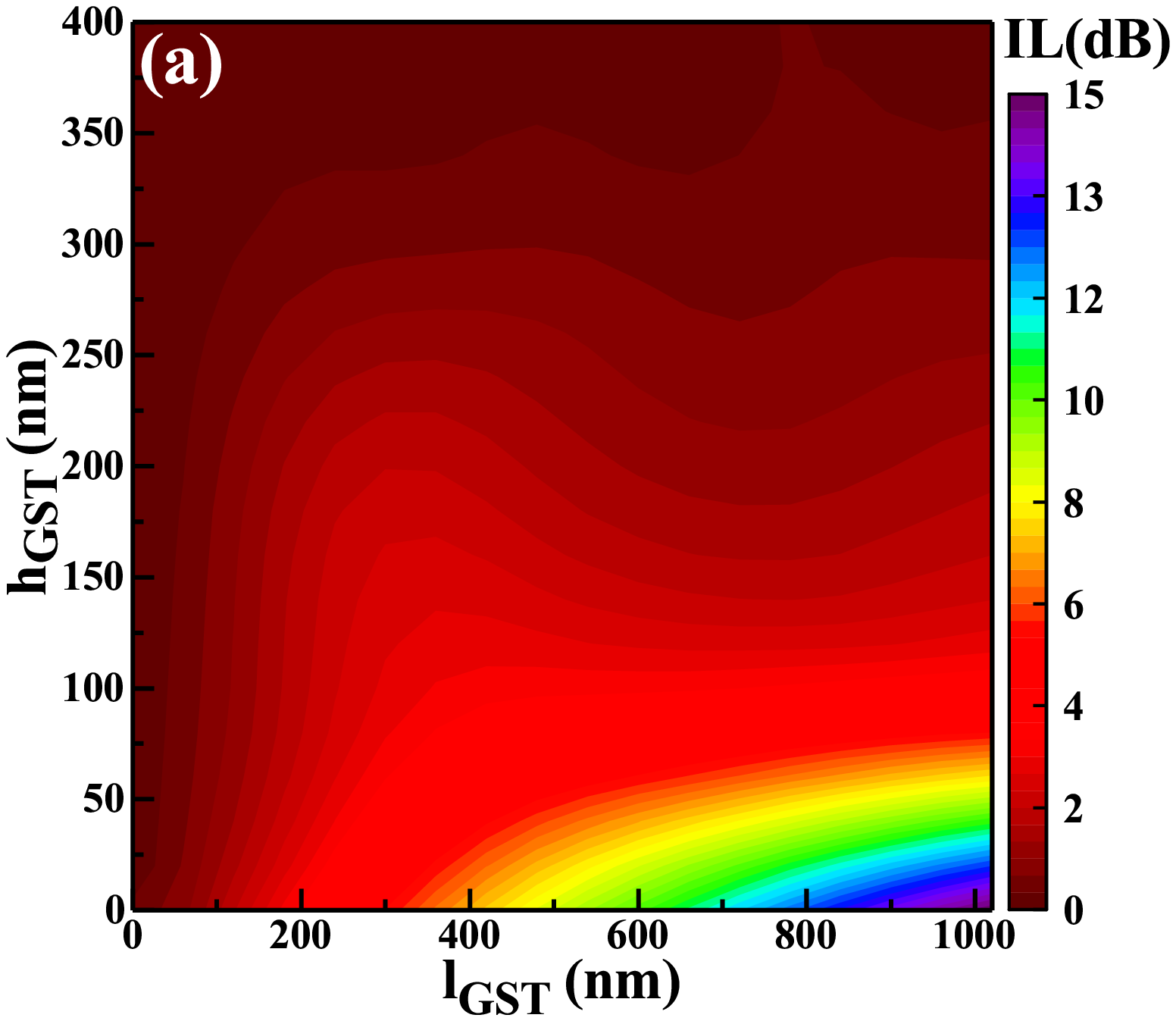}}\hspace{0.1em}
%\label{fig_first_case}
%\vfil
\subfloat{\includegraphics[scale=0.5]{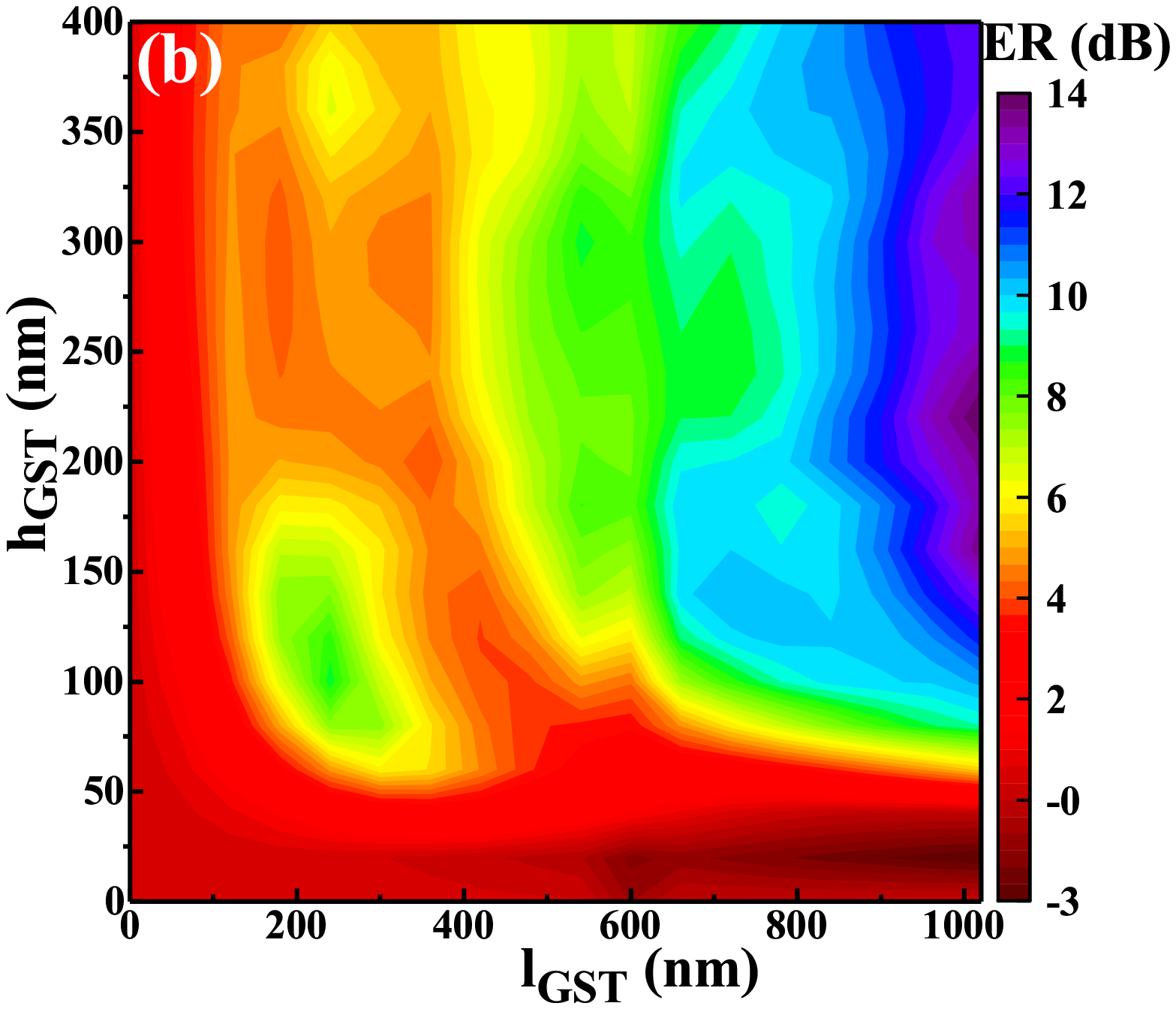}}
%\label{fig_second_case}
\caption{Contour plots of (a) IL and (b) ER as a function of $h_{GST}$ and $l_{GST}$ for 1 $\times$ 1 waveguide switch with GST substituted in fully etched silicon waveguide.}
\label{fig4}
\end{figure}
\begin{figure}
	\centering
		\includegraphics[scale=.5]{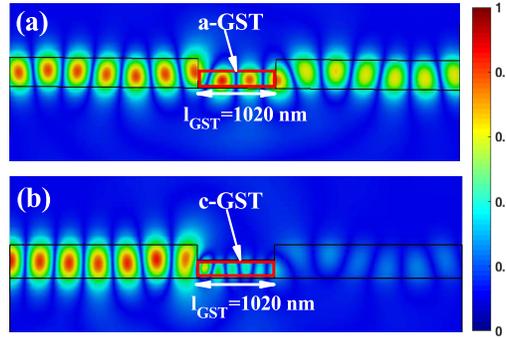}
	\caption{Three-dimensional simulations of the designed hybrid Si-GST 1 $\times$ 1 waveguide switch with optimized values of GST dimensions substituted in fully etched silicon waveguide for the (a) ON state (a-GST) and the (b) OFF state (c-GST).}
	\label{fig5}
\end{figure}
In the second design of 1 $\times$ 1 waveguide switch, the GST is embedded in partially etched Si waveguide as shown in figure~\ref{fig2}(a). This approach is taken to minimize the interface scattering and reflection losses between the active waveguide region and the adjacent passive waveguide regions. This approach drastically reduce insertion loss of the device. For optimization, we varied the $l_{GST}$, $h_{GST}$ and $h_{Si}$, where $h_{Si}$ is the height of the partially etched Si as shown in figure~\ref{fig2}(b). Both $h_{GST}$ and $h_{Si}$ are varied in such a way that the total height of the waveguide ($h_{GST}$ + $h_{Si}$ = 400 nm) remains the same. The mode profile of switch active region for the ON and OFF states is shown in figure~\ref{fig2}(c).\\
For both of the switch designs, we calculated switch parameters insertion loss (IL) and extinction ratio (ER). Both IL and ER are calculated from the switch transmission data. These are defined as IL = 10log($T_{ON}$), and ER = 10log($T_{ON}$/$T_{OFF}$), where $T_{ON}$ and $T_{OFF}$ denotes transmission for the amorphous and crystalline phase respectively. Based on the best value of ER and IL, we selected the dimensions for the electro-thermal simulations. 
\subsection{1 $\times$ 2 directional coupler switch}
Figure~\ref{fig3} depicts the parallel waveguides region of the proposed 1 $\times$ 2 directional coupler switch. For simplicity only the parallel waveguides are shown in figure~\ref{fig3}(a). To achieve controlled coupling of optical mode from one waveguide to other, a 20 nm thick layer of GST is embedded in partially etched Si waveguide as shown in the cross section view of active waveguide region in figure~\ref{fig3}(b). This GST thickness is selected to minimize the phase mismatch between the Si and the hybrid Si-GST waveguides. The gap between the coupling waveguides is denoted as g, and the length of the active region of the coupler is denoted as $l_{GST}$. As in the previous case, the ITO electrodes are used to voltage trigger phase change of GST. We calculated the various parameters of directional coupler switch such as coupling ratio (CR) = $P_{cross}$/($P_{cross}$+$P_{bar}$); excess loss (EL) = 10log$P_{i}$/($P_{cross}$ + $P_{bar}$); and insertion losses IL, $IL_{cross}$ = 10log($P_{i}$/$P_{cross}$) and $IL_{bar}$ = 10log($P_{i}$/$P_{bar}$), where $P_{i}$ denotes the input power, $P_{bar}$ and $P_{cross}$ denotes the output power for the bar and the cross state of switch, respectively. By taking into consideration the value of CR and IL, we selected the optimized value of g and performed the electro-thermal simulations of switches.\\
All the simulations are performed using CST-Microwave Studio. The electromagnetic investigations are done using three dimensional finite integration technique with open boundary conditions. The thermal investigations are performed by electro-thermal co-simulations using Mphysics module of this software. For the simulations performed, we define the refractive index of GST as n + ik, where n and k are the real and imaginary refractive index of GST. The refractive indices used are 4.05 + i0.006 (a-GST) and 6.80 + i0.40 (c-GST) at 2.1 $\mu$m wavelength. The ITO, $SiO_{2}$, Si and N-Si refractive indices are set to 1.84 + i0.019, 1.45, 3.45, 3.45 + i0.0003 respectively \cite{b19,b21}.
\begin{figure}
\centering
\subfloat{\includegraphics[scale=0.5]{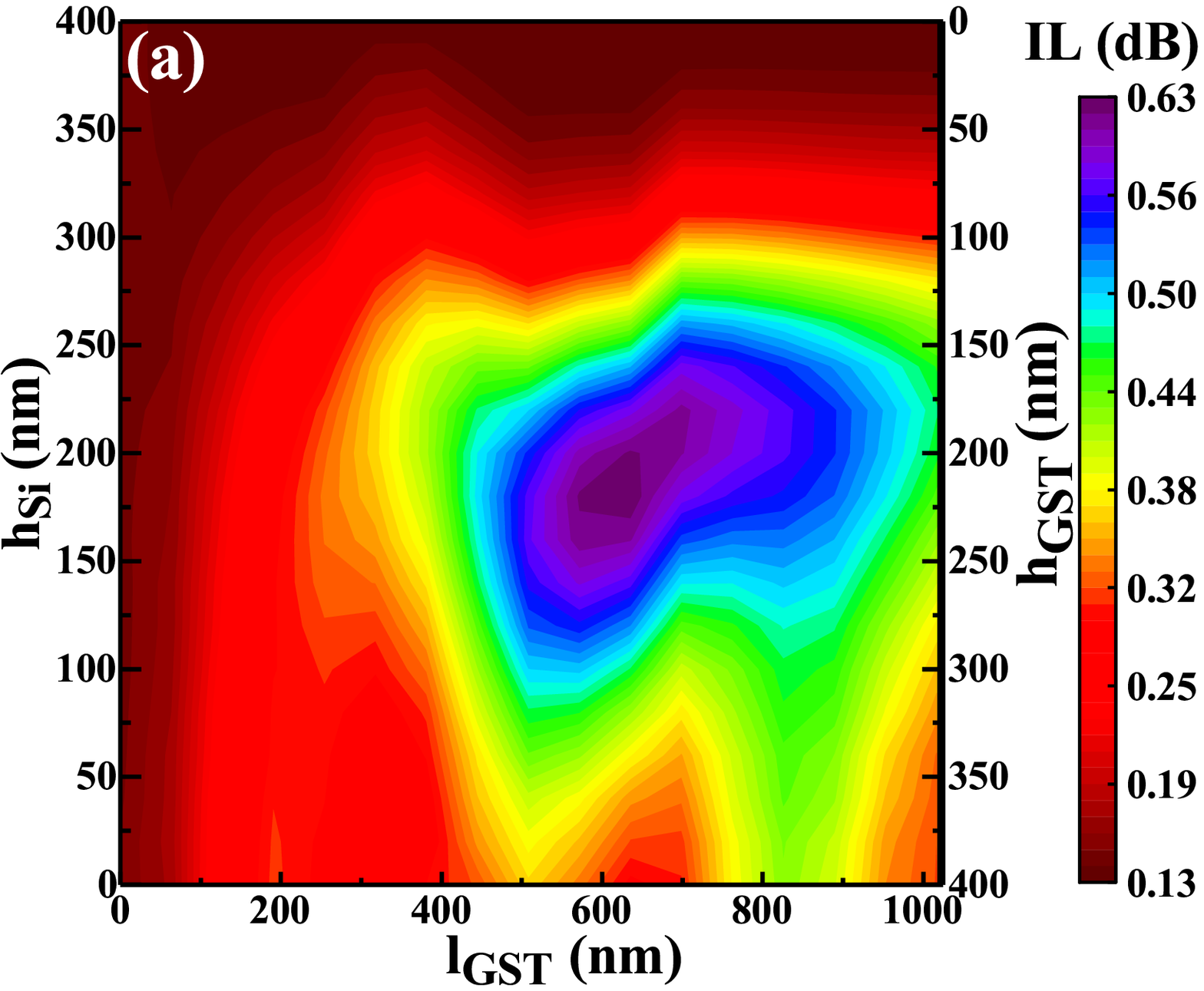}}
%\label{fig_first_case}
%\vfil
\subfloat{\includegraphics[scale=0.5]{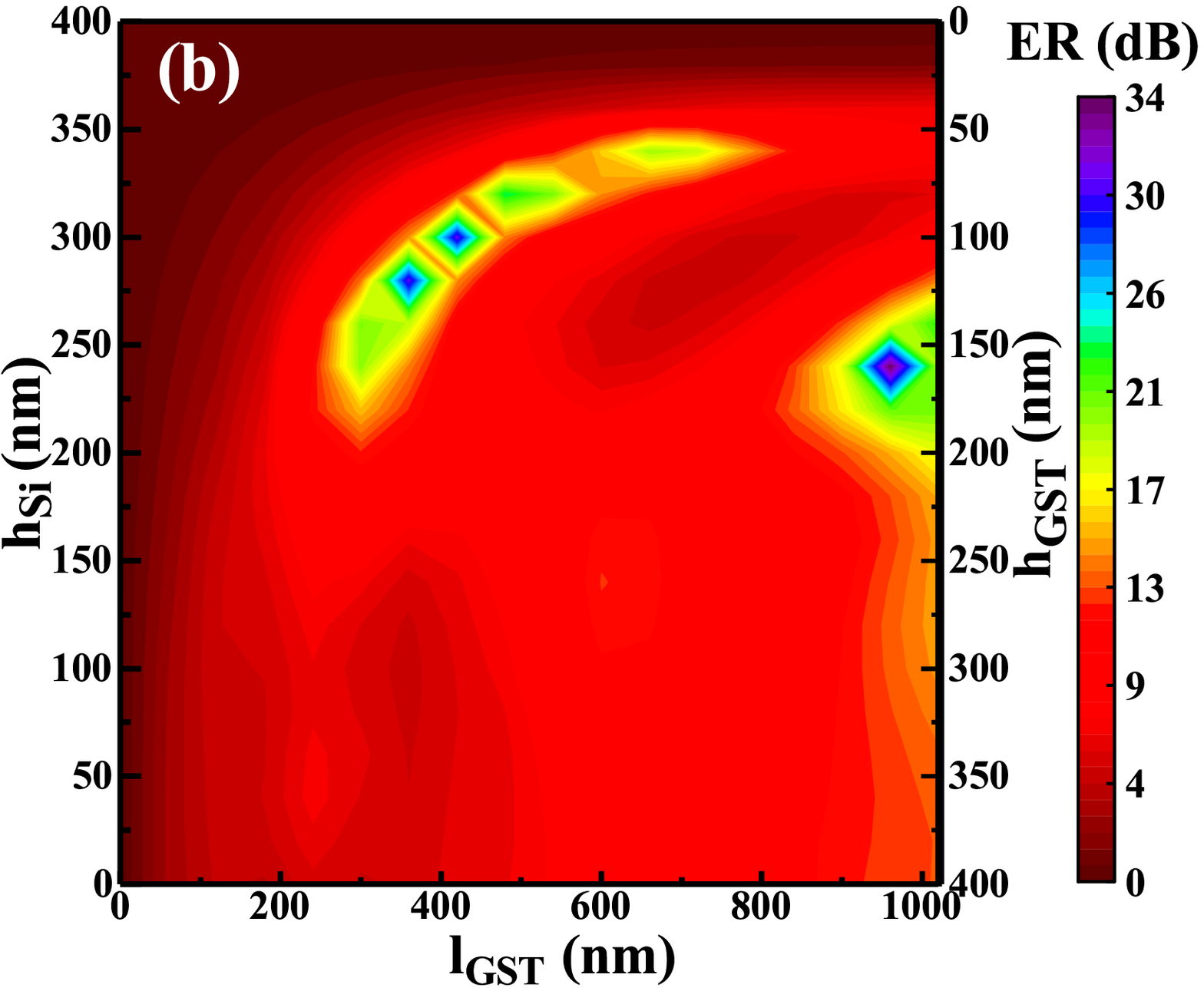}}
%\label{fig_second_case}
\caption{Contour plots of (a) IL and (b) ER as a function of $h_{GST}$, $h_{Si}$ and $l_{GST}$ for 1 $\times$ 1 waveguide switch with GST substituted in the partially etched silicon waveguide.}
\label{fig6}
\end{figure}
\begin{figure} [t]
	\centering
		\includegraphics[scale=.5]{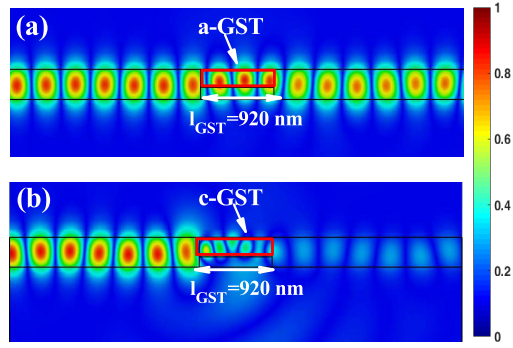}
	\caption{Three-dimensional simulations of the designed hybrid 1 $\times$ 1 Si-GST waveguide switch with optimized values of GST dimensions substituted in partially etched silicon waveguide for the (a) ON state (a-GST) and the (b) OFF state (c-GST).}
	\label{fig7}
\end{figure}

%%%%%%%%%%%%%%%%%%%%%%%%%%%%%%%%%%%%%%%%%% Results %%%%%%%%%%%%%%%%%%%%%%%%%%%%%%%%%
%%%%%%%%%%%%%%%%%%%%%%%%%%%%%%%%%%%%%%%%%%%%%%%%%%%%%%%%%%%%%%%%%%%%%%%%%%%%%%%%%%%%%
\section{Analysis and results}
\subsection{Electromagnetic analysis}
For 1 $\times$ 1 waveguide switch depicted in figure~\ref{fig1}, the contour plots of IL and ER as a function of $l_{GST}$ and $h_{GST}$ are shown in figure~\ref{fig4}. With the increase in $l_{GST}$, both IL and ER increase for $l_{GST}$ \textless 1 $\mu$m. For any fixed value of $l_{GST}$, the decrease in $h_{GST}$ results in increased IL due to low confinement of optical mode in thinner guiding layer. Most part of the optical mode sees the air as it reaches the etched region of the waveguide and hence the increased scattering and reflection losses. As can be seen from figure~\ref{fig4}(a); for $h_{GST}$ \textless 80 nm and $l_{GST}$ \textgreater 320 nm, switch exhibits high IL in the range of 6--15 dB. As shown in figure~\ref{fig4}(b), the negative ER is observed in this region since the loss in a-GST is larger than c-GST. This happens due to lower confinement of optical mode in a-GST as compared to that in c-GST owing to higher real part of refractive index of c-GST in comparison to that of a-GST. With the increase in $h_{GST}$ beyond 80 nm, the ER continues to increase, but saturates to around $\sim$ 14 dB. Figure~\ref{fig5} shows the electric field profile along the length of the waveguide. In this figure the value of $h_{GST}$ and $l_{GST}$ is 240 nm and 1020 nm respectively. For these dimensions the IL and the ER are found to be 1.36 dB and 14 dB respectively. The back reflection in this case is -11.98 dB and -26.73 dB for ON and OFF state respectively. For obtaining the improved performance, we investigated the waveguide switch as shown in figure~\ref{fig2}. This design gives drastic improvement in switch performance in terms of IL and ER and at the same time provides reduced scattering and back reflections.
\begin{figure}
\centering
\subfloat{\includegraphics[scale=0.5]{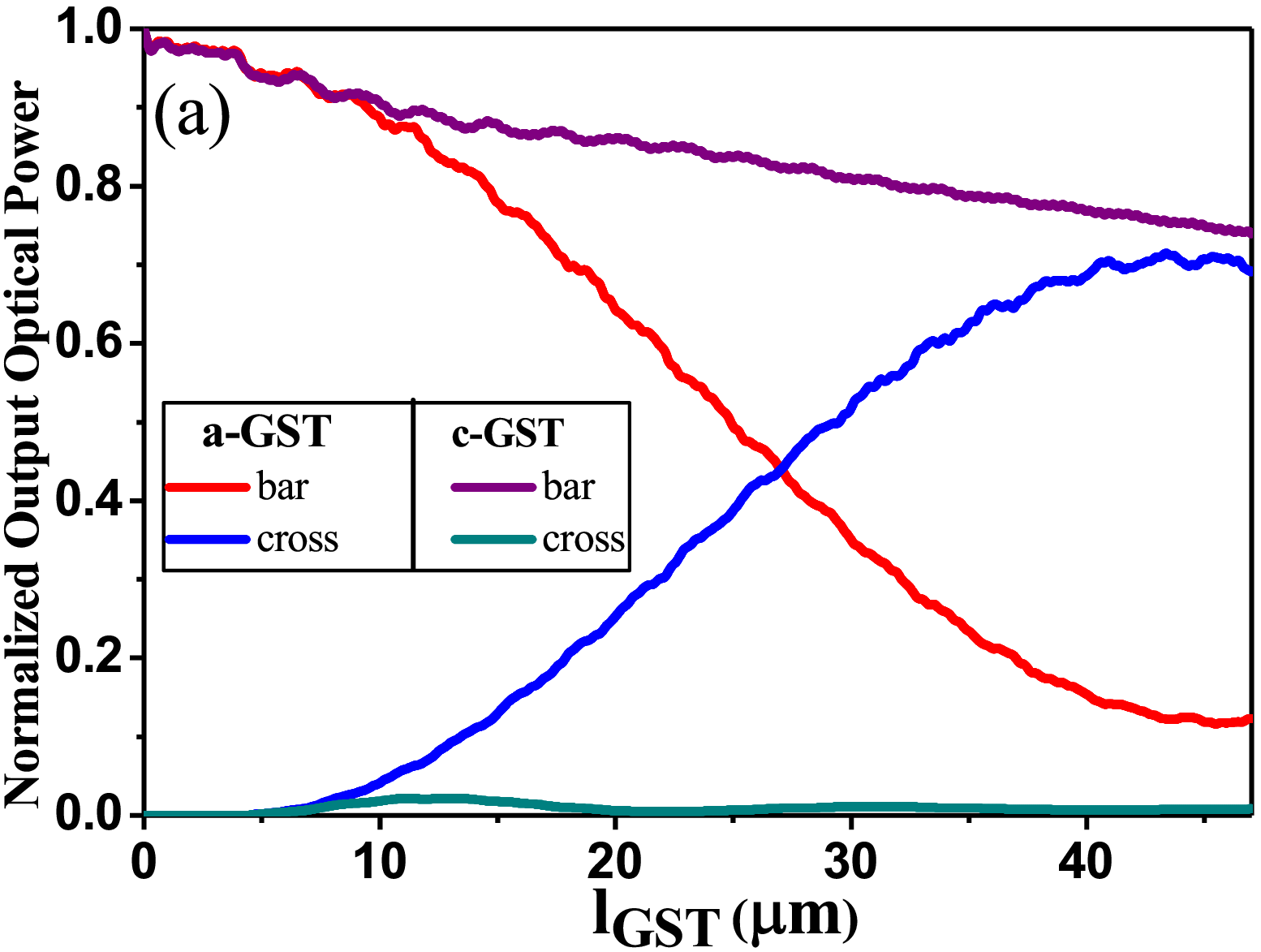}}\hspace{0.1em}
%\label{fig_first_case}
%\vfil
\subfloat{\includegraphics[scale=0.55]{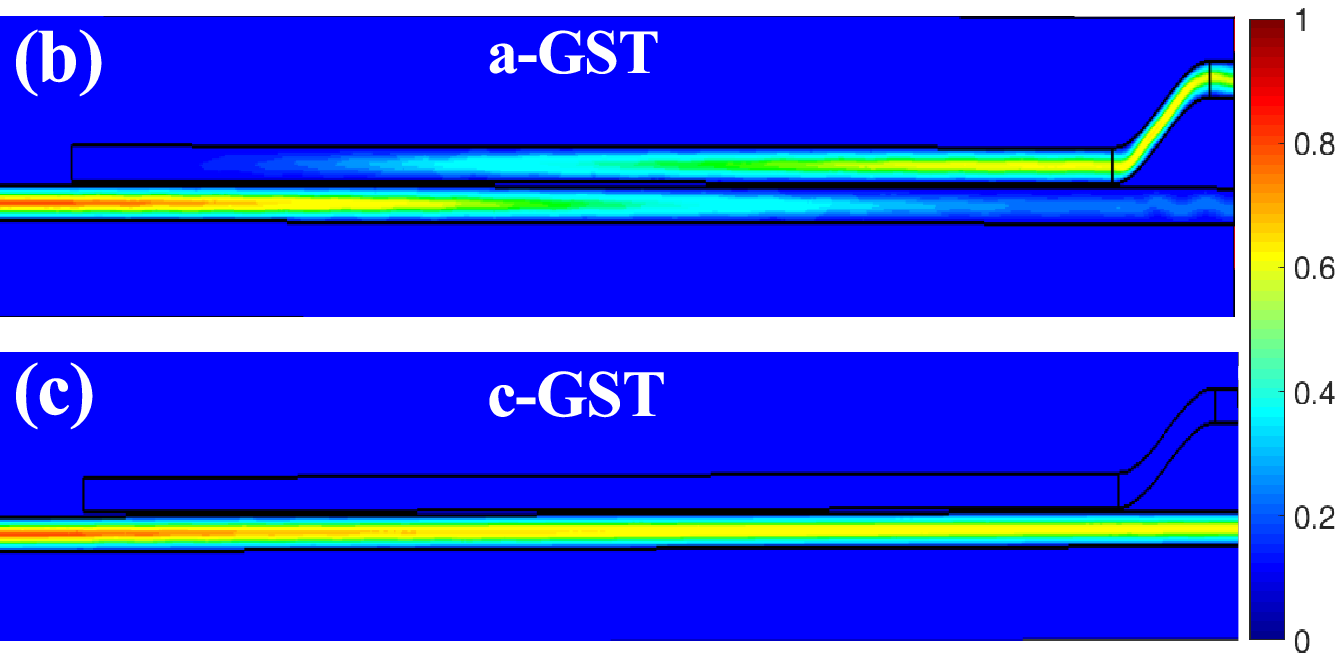}}
%\label{fig_second_case}
\caption{(a) Normalized output optical power as a function of device length $l_{GST}$ for the cross and the bar state. The power distribution in (b) cross state and (c) bar state.}
\label{fig8}
\end{figure}
For this switch, the contour plots of IL and ER are shown in figure~\ref{fig6}. The maximum ER obtained is 34.04 dB with an IL of 0.49 dB. These values of ER and IL are obtained for the GST volume of 800 nm $\times$ 260 nm $\times$ 920 nm (w $\times$ h $\times$ l). The back reflection for this design is -16.71 dB and -28.73 dB for ON and OFF state respectively. When compared with previous design there is an improvement of 14 dB in IL and 20 dB in ER. The optical field profiles for the optimized switch in the ON and OFF states are shown in figure~\ref{fig7}.\\
\begin{table*}[!t]
\renewcommand{\arraystretch}{1.3}
% if using array.sty, it might be a good idea to tweak the value of
%\extrarowheight as needed to properly center the text within the cells
\caption{Calculated Parameters of 1 $\times$ 2 Directional Coupler Switch With Amorphous and Crystalline Phases of GST.}
\label{table1}
\centering
\begin{tabular}{|c|c|cccc|cccc|}
   \hline
     Gap &  Coupling length &  & Amorphous& &  &  &Crystalline& & \\ 
     \hline
    g & $l_{c}$  & CR  & $EL$ & $IL_{cross} $ & $IL_{bar}$ &CR &$ EL$ & $ IL_{cross}$& $IL_{bar}$ \\
    (nm)& ($\mu$m) & \% &(dB)&(dB) &(dB) &\% &(dB) &(dB)&  (dB)\\
   \hline
   50  & 36 & 92.00 &1.07&1.77 &9.83 &1.04 &1.13 &20.02&  1.18\\
  75  & 42 &91.64 &1.08&1.52 & 11.92&1.60 &1.23 &18.95&  1.33\\
   100 & 52 & 85.56&1.1&1.90  &9.66  &1.10 &1.13 &20.49&  1.33\\
   125 & 56 & 79.78&1.22&2.24 & 8.18 &0.83 &1.18 &22.06&  1.28\\
   150 & 65 &73.88 &1.41&2.72 &7.22  &0.57 &1.39 &23.91&  1.37\\
    175 & 73 &57.14 &1.54&3.95 & 5.10 &0.27 &1.41 &26.75&  1.39\\
  200 & 82 &47.05 &1.66&4.76 & 4.42 &0.19 &1.53 &28.43&  1.48\\
   \hline
  \end{tabular}
  \end{table*}

In 1 $\times$ 2 directional coupler switch, we plotted output optical power emerging from the bar and the cross port as a function of device length $l_{GST}$. Using these plots, coupling length $l_{c}$ are obtained for different values of coupler gap 'g'. To give a particular example, the normalized optical power with the value of g = 100 nm for both phases of the GST is plotted in figure~\ref{fig8}(a). In case of a-GST, the light launched from the Si waveguide couples to the hybrid waveguide after traveling a distance equal to the coupling length $l_{c}$ = 52 $\mu$m with a coupling ratio of 85.56 \%. However, in case of c-GST, the coupling is weaker due to large difference in refractive index between Si and hybrid waveguide, therefore almost all the light remains in the Si waveguide even after travelling a distance equal to $l_{c}$ and the coupling ratio is just 1\%. The high output can be obtained in cross or bar state by selection of proper value of $l_{GST}$. The calculated performance parameters of the switch are listed in Table \ref{table1}. As can be seen from the table that in both phases of GST, the CR decreases with increase in the value of g. This is due to the dependence of power transmission along the device length on the optical phase matching. The switch corresponding to the value of g = 100 nm can provide ER of 18.59 dB and 8.33 dB in cross and bar state respectively. For such an extinction ratio the active length required is 52 $\mu$m. The cross and bar states of this switch for g value of 100 nm are illustrated in figure~\ref{fig8}(b) and (c) respectively.
%%%%%%%%%%%%%%%%%%%%%%%%%%%%%%%%%%%%% Thermal results  %%%%%%%%%%%%%%%%%%%%%%%%%%%%%%%%%%%%%%%%%%%%%%%%
%%%%%%%%%%%%%%%%%%%%%%%%%%%%%%%%%%%%%%%%%%%%%%%%%%%%%%%%%%%%%%%%%%%%%%%%%%%%%%%%%%%%%%%%%%%%%%%%%%%%%%%
\begin{figure}
	\centering
		\includegraphics[scale=.6]{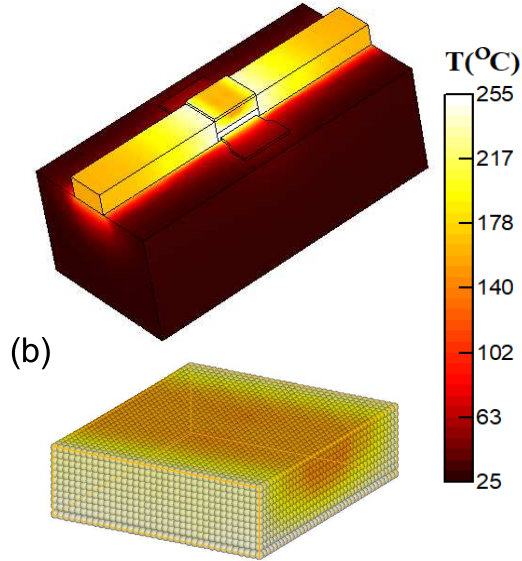}
	\caption{(a) The temperature profile of optimized 1 $\times$ 1 waveguide switch and (b) zoom in view of GST region for the process of crystallization.}
	\label{fig9}
\end{figure}
\begin{figure}[t]
	\centering
		\includegraphics[scale=.6]{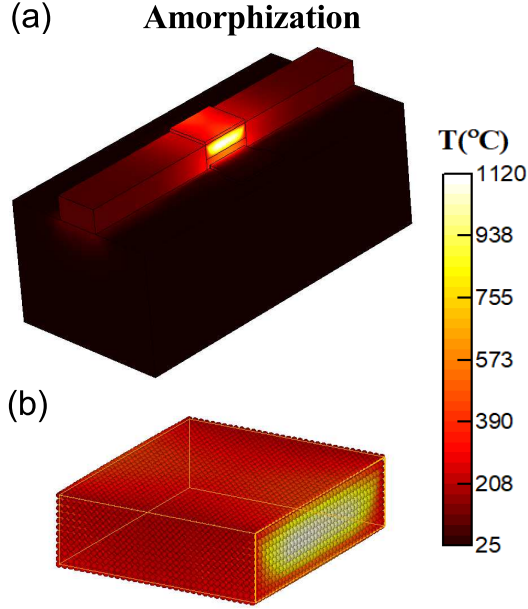}
	\caption{(a) The temperature profile of optimized 1 $\times$ 1 waveguide switch and (b) zoom in view of GST region for the process of amorphization.}
	\label{fig10}
\end{figure}
\subsection{Thermal analysis} 
In this section, we present the investigations of electrically induced phase transition in GST using voltage pulses. The phase transition of GST from a-GST to c-GST and vice-versa is induced by heating the GST through Joule heating. For this purpose, we implemented an electro-thermal co-simulation model using the CST Microwave Studio. In this co-simulation model, the electrical simulations predict the current and voltage distributions in the device, while thermal simulations are used to obtain temperature distribution in the device. \\
The governing equations for the electrical and thermal simulations are the following:
\begin{eqnarray}
\nabla \cdot [\sigma (x,y,z,t) \nabla V] = 0\\
\nabla \cdot [k(x,y,z) \nabla T] + Q = C_v (\frac{dT}{dt}) \\
Q = \sigma E^{2}
\end{eqnarray}
In (1), $\sigma$ and $V$ represents the electrical conductivity of the material and applied voltage respectively. In (2), $k$, $T$ and $C_v$ denotes the thermal conductivity, temperature and volumetric heat capacity respectively while Q is the Joule heat generation associated with electric field (E) and is represented by (3). For GST, the electrical conductivity depends upon temperature and its phase, and exhibit a sharp increase in conductivity on transformation from a-GST ($\sigma_{a-GST} = 3 {\Omega}^{-1} m^{-1} $) to c-GST ($\sigma_{c-GST} = 2770$ ${\Omega}^{-1} m^{-1} $) \cite{b22}. Material properties used for the thermal simulations are presented in Table \ref{table2} \cite{b23,b24,b25}. \\
\begin{table}[!b]
%% increase table row spacing, adjust to taste
\renewcommand{\arraystretch}{1.3}
% if using array.sty, it might be a good idea to tweak the value of
%\extrarowheight as needed to properly center the text within the cells
\caption{Coefficients Used for the Thermal Simulations}
\label{table2}
\centering
%% Some packages, such as MDW tools, offer better commands for making tables
%% than the plain LaTeX2e tabular which is used here.
\begin{tabular}{|c|c|c|c|c|c|}
\hline
Coefficients & $SiO_{2}$ & Si & ITO & a-GST & c-GST\\
\hline
Density  (Kg/$m^{3}$) & 2202 & 2329 & 6800 & 6150 & 6150 \\
Specific heat  (J/Kg K)& 746 & 713 & 1340 & 212 & 212 \\
Thermal cond.  (W/mK) & 1.38 & 140 & 5 & 0.19 & 1.58 \\
\hline
\end{tabular}
\end{table}
In order to find the voltage and energy required for complete phase transformation of the GST, we recorded the rise in temperature for a different set of applied voltages. The duration of voltage pulses is chosen as 100 ns and 10 ns for amorphous to crystalline and crystalline to amorphous phase transformation respectively \cite{b26,b27}.
For crystallization, the GST layer should be heated above 140 $^{\circ}$C but below 546 $^{\circ}$C (melting temperature) to induce complete phase transition. For 1 $\times$ 1 waveguide switch, results are presented for the optimized dimensions of switch design, in which an 0.92 $\mu$m length GST layer having a cross section of 800 $\times$ 260 $nm^{2}$ is embedded in partially etched Si waveguide, as shown in figure \ref{fig7}. The Si layer beneath the GST is N-type doped with doping density of $10^{-17}$ $cm^{-3}$ to provide adequate electrical conductivity ($\sigma = 10^{3} \Omega^{-1} m^{-1} $). The doping of Si layer leads to the slight modification in IL and ER. The new values of IL and ER are 0.52 dB and 33.79 dB respectively. Our simulation results show that switching to the crystalline phase can be achieved by applying a 5 V rectangular voltage pulse of 100 ns duration, which increase the maximum temperature of the GST to 255 $^{\circ}$C. Figure \ref{fig9}(a) shows the spatial distribution of temperature in the device for the process of crystallization. To confirm whether the whole GST volume temperature is raised above the 140 $^{\circ}$C, we analyzed three dimensional thermal profile of the GST region illustrated in Fig \ref{fig9}(b). The zoom-in view of GST indicates that the crystallization is induced in the whole GST layer. The energy consumed during the process is estimated to 0.9 nJ.\\ To induce amorphization, the GST material should be heated above the melting temperature for a short amount of time, and cools down to the disordered amorphous phase. Our simulation results demonstrate that a voltage pulse of 7.5 V applied for 10 ns duration increased the device temperature to a maximum of 1120 $^{\circ}$C. The energy consumed corresponding to the applied voltage is 22.8 nJ. Spatial distribution of temperature for the process of amorphization is shown is figure \ref{fig10}(a). The zoom-in view of GST layer shown in Fig \ref{fig10}(b) confirms that the amorphization is induced in the whole GST layer for the considered voltage.\\
\begin{figure}
	\centering
		\includegraphics[scale=.6]{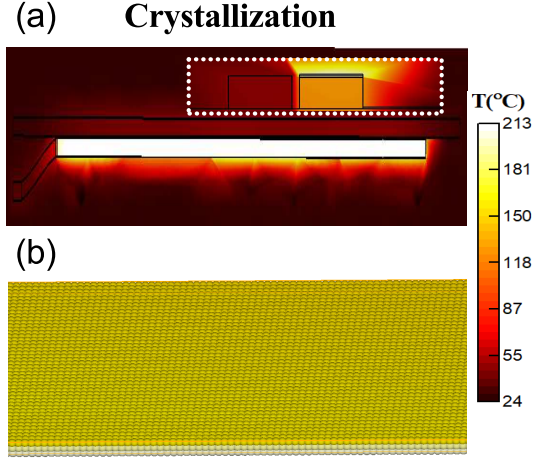}
	\caption{(a) The temperature profile of optimized 1 $\times$ 2 directional coupler switch and (b) zoom in view of center of GST loaded arm for the process of crystallization.}
	\label{fig11}
\end{figure}
\begin{figure}
	\centering
		\includegraphics[scale=.6]{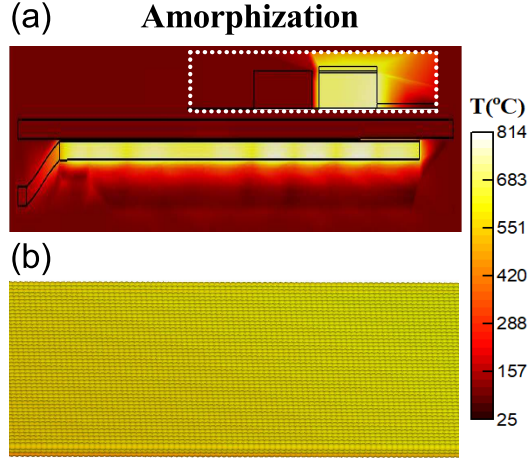}
	\caption{(a) The temperature profile of optimized 1 $\times$ 2 directional coupler switch and (b) zoom in view of center of GST loaded arm for the process of amorphization.}
	\label{fig12}
\end{figure}
The above mentioned procedure is also used for the thermal analysis of 1 $\times$ 2 directional coupler switch with a gap of 100 nm and coupling length of 52 $\mu$m. Like the case of 1 $\times$ 1 switch, the Si layer beneath the GST is doped with N-type impurity. Such doping of silicon layer results in an ER of 10.33 dB and 5.23 dB for the cross and bar state respectively. For crystalline to amorphous phase transition, we applied a 6 V voltage pulse for a duration 100 ns. This pulse raised the temperature of GST region to a maximum of 213 $^{\circ}$C as can be seen from the temperature profile of switch. The top view of the same is shown in figure \ref{fig11}(a) while inset shows the cross-section view. The zoom in view from the center of GST loaded arm is shown in figure \ref{fig11}(b). As can be seen from figure \ref{fig11}(b) the temperature rise is sufficient for crystallization of whole GST region. The energy consumed in this process is 5.77 nJ. For the reverse process, a voltage pulse of 7.5 V of duration 10 ns raised the maximum temperature of the GST to 814 $^{\circ}$C as can be seen from the temperature profile of the switch. The top view of the same is shown in figure \ref{fig12}(a) while inset shows the cross-section view. The zoom in view from the center of the GST loaded arm is shown in figure \ref{fig12}(b). As can  be seen from the figure \ref{fig12}(a) the temperature rise is sufficient for amorphization of whole GST region. The energy consumption for the process of amorphization is 38.9 nJ.

%%%%%%%%%%%%%%%%%%%%%%%%% Conculsions %%%%%%%%%%%%%%%%%%%%%%%%
%%%%%%%%%%%%%%%%%%%%%%%%%%%%%%%%%%%%%%%%%%%%%%%%%%%%%
\section{Conclusion}
We have designed and theoretically analyzed ultra-compact hybrid Si-GST switches for the emerging wavelengths window of Mid-IR around 2 $\mu$m. In a 1 $\times$ 1 waveguide switch, we obtained 33.79 dB extinction ratio with 0.52 dB insertion loss at 2.1 $\mu$m for an optimized GST length of only 0.92 $\mu$m. Reversible on-off switching is achieved by electrical actuation of GST through ITO electrodes. The amorphous to crystalline transition in GST is induced by a 5 V voltage pulse corresponding to the energy consumption of 0.9 nJ. While in case of crystalline to amorphous transition a 7.5 V voltage pulse is sufficient to increase the GST temperature to 1120 $^{\circ}$C, and the energy consumed during this process in 22.8 nJ. Moreover in both cases, the phase change happens in the complete GST region. In a 1 $\times$ 2 directional coupler switch, we showed that reversible switching between the cross-bar state can be achieved with an extinction ratio of 10.33 dB and 5.23 dB in the cross and bar state respectively with an optimized active length of only 52 $\mu$m. The voltages required for amorphization and crystallization of GST are 6 V and 7.5 V respectively. The energy consumed for a complete cycle of phase transition is 44.67 nJ. Our work will help in experimental realization of ultra-compact high on-off ratio electro-optic switches for inter-chip and intra-chip photonic applications.
\section{Acknowledgements}
This work is supported by the research projects funded by the IIT Roorkee under FIG scheme; and Science and Engineering Research Board (SERB), Department of Science and Technology, Govt. of India under Early Career Research award scheme (project File no. ECR/2016/001350)
\bibliographystyle{unsrt}

\end{document}